%% file: sample-authordraft.tex
\documentclass[sigconf]{acmart}
\usepackage{booktabs}
\usepackage{threeparttable}
\AtBeginDocument{%
  \providecommand\BibTeX{{%
    \normalfont B\kern-0.5em{\scshape i\kern-0.25em b}\kern-0.8em\TeX}}}

\setcopyright{rightsretained}
\copyrightyear{2021}
\acmYear{2021}
\acmDOI{}
\acmISBN{}

\acmConference[KDD 2nd Workshop on Data-driven Humanitarian Mapping '21]{Singapore '21: ACM SIGKDD Conference on Knowledge Discovery and Data Mining}{August 14--18, 2021}{Singapore}

\begin{document}

\title{Under the Radar - Auditing Fairness in ML for Humanitarian Mapping}

\author{Lukas Kondmann}
\email{lukas.kondmann@dlr.de}
\orcid{1234-5678-9012}
\affiliation{%
  \institution{German Aerospace Center (DLR) \& Technical University of Munich (TUM)}
  \city{Munich}
  \country{Germany}
}

\author{Xiao Xiang Zhu}
\email{xiaoxiang.zhu@dlr.de}
\orcid{1234-5678-9012}
\affiliation{%
  \institution{German Aerospace Center (DLR) \& Technical University of Munich (TUM)}
  \city{Munich}
  \country{Germany}
}

\renewcommand{\shortauthors}{Kondmann and Zhu}

\begin{abstract}
Humanitarian mapping from space with machine learning helps policy-makers to timely and accurately identify people in need. However, recent concerns around fairness and transparency of algorithmic decision-making are a significant obstacle for applying these methods in practice. In this paper, we study if humanitarian mapping approaches from space are prone to bias in their predictions. We map village-level poverty and electricity rates in India based on nighttime lights (NTLs) with linear regression and random forest and analyze if the predictions systematically show prejudice against scheduled caste or tribe communities. To achieve this, we design a causal approach to measure counterfactual fairness based on propensity score matching. This allows to compare villages within a community of interest to synthetic counterfactuals. Our findings indicate that poverty is systematically overestimated and electricity systematically underestimated for scheduled tribes in comparison to a synthetic counterfactual group of villages. The effects have the opposite direction for scheduled castes where poverty is underestimated and electrification overestimated. These results are a warning sign for a variety of applications in humanitarian mapping where fairness issues would compromise policy goals. 
\end{abstract}




\maketitle

\section{Introduction}
Satellite images have become a widely used data source across disciplines to map humanitarian indicators in low- and middle-income countries. Earth observation data are, for example, key to assessing damages through natural disasters \cite{gupta2019creating,moya2020detecting}, crop yields \cite{lobell2015scalable} urbanization, \cite{elvidge1997mapping}, as well as economic development \cite{chen2011using}. Among the most frequently used input for predicting local levels of development is light activity at night observed from space \cite{bennett2017advances}. Nighttime lights (NTLs) are, among others, closely related to population \cite{sutton1997modeling}, electricity \cite{elvidge1997mapping}, and local levels of poverty \cite{bruederle2018nighttime}. More recently, daytime and nighttime satellite images have been increasingly used in conjunction with machine learning which further improved the accuracy of spatial predictions of local economic development \cite{jean2016combining,yeh2020using}. Recent evidence, however, suggests that algorithmic predictor systems
can also deepen existing inequalities or biases \cite{o2016weapons} which is particularly relevant for applications close to policy-making.  

Following these concerns, fairness audits of algorithmic decision systems have been pioneered in a variety of fields in the Global North but potential blind spots in the low- and middle-income countries are less in the scope of current literature  \cite{noriega2020algorithmic}. Noriega-Campero et al. \cite{noriega2020algorithmic} were among the first to study algorithmic fairness in policy-making in the developing world but focus on targeting poor households based on survey data. 

The success of humanitarian mapping with big data has, however, equipped policy-makers with an accurate, low-cost, and frequent alternative to surveys. For example, Togo has recently relied on a combination of satellite images and mobile phone metadata to target households for the allocation of Covid-19 relief funds. \footnote{https://www.bbc.com/news/av/stories-56580833} The use of poverty mapping systems in practice also requires the application of fairness guidelines \cite{burke2021using}. However, there is currently little to no evidence if humanitarian mapping from space is prone to fairness issues.

In this paper, we study if a set of current methods in humanitarian mapping shows prejudice for or against a sample of marginalized communities. We use the setting of India by analyzing the case of discrimination against scheduled castes and tribes which has been a topic of societal and political relevance for decades  \cite{banerjee1985caste}. We map electricity and poverty on the village level in India from NTLs with a linear regression and random forests. Then, we audit if prediction errors are systematically related to scheduled caste and tribe population. For this purpose, we match villages of interest with synthetic counterfactual villages to identify a potential bias in predictions between the two groups. Our fairness identification strategy is inspired by propensity score matching and contributes to the growing field of causal inference-based approaches in fairness audits \cite{kilbertus2017avoiding}.

We find a significant bias for both scheduled tribes and castes that is notable in magnitude and similar for both methods used. When mapping poverty in villages with scheduled tribe communities, predicted poverty rates are on average over one percentage point higher than in the comparison group. For scheduled castes, the effect goes in the opposite directions and is slightly smaller in magnitude. These results are a strong indication that geospatial data in AI applications may still be subject to fairness issues and a more thorough understanding of these issues is necessary. To the best of our knowledge, this is the first study to audit fairness in humanitarian mapping applications from space.  

\section{Historical Background of Castes and Tribes in India}
Historically, Indian society was separated into a hierarchy of subgroups (`castes') originating from the Hindu system of 'varna' \cite{banerjee2009labor}. The origins and the contemporary role of the caste system in India are a complex topic of anthropological research \cite{beteille1992backward,srinivas1957caste}.
However, there is a well-understood notion of the top and bottom groups in the historical caste hierarchy \cite{banerjee2009labor}. The highest-ranked group are the Brahmin castes which, together with other groups, constitute the `upper castes'. At the bottom of the caste system, a large subgroup was treated as `untouchable' by the upper class with explicit segregation policies such as restricted access to public buildings or education. 

In 1950, the constitution declared discrimination against these groups, so called `scheduled castes' which made up about 16.2\% of the population in 2001, illegal. A variety of affirmative action policies were implemented to support their economic and social catch-up. Similarly, India's endogenous tribe population was also included in these policies since they the lowest educational and social indicators in all of India's social groups \cite{banerjee2009labor}. While there has been a notable increase in education and economic standing in these marginalized communities, overall levels remain low compared to the national average. Further, caste and tribe association still matter in contemporary Indian society. Particularly, scheduled caste and tribe association may be linked to discrimination in the labor market \cite{banerjee1985caste}. Further, preferences about the caste of potential partners in arranged marriages suggest some form of perceived cultural persistence of the historical hierarchy \cite{ahuja2016crossing}.

Therefore, the role of scheduled castes and tribes are still a present and important issue today.  Beyond Indian policy, they serve as an example for a historically marginalized community who may face significant hurdles because of their subgroup association. We study their case because of the longstanding efforts to combat discrimination against them by Indian policy and the societal relevance of caste association.

\section{Data}
We obtain village-level development characteristics from the Socio-Economic High-Resolution Rural-Urban Geographic Dataset on India (SHRUG) \cite{asher2019socioeconomic}. The SHRUG combines village data from Indian censuses for virtually all Indian villages over the timespan from 1991-2011 with consistent geographic units over time. This ensures a stable unit of analysis because different villages may be, for example, joined together over time. We focus our analysis on the year 2011 for poverty since this is the only year where consumption information was collected. For electricity, we use census data and NTLs from 2001. Additionally, we include the geospatial village coordinates from the NASA SEDAC India Village-Level Geospatial Socio-Economic Data Set. Table \ref{tab:Summary} presents the summary statistics of key variables in the dataset. 

\begin{table*}[!htbp]
\centering
\caption{Summary Statistics}
\begin{threeparttable}
\footnotesize
\setlength{\tabcolsep}{1.3\tabcolsep}
\input{summary}\textbf{}
\end{threeparttable}
\label{tab:Summary}
\end{table*}

In total, we restrict our analysis to 525,255 villages. This represents almost the entirety of villages in India with only a small subset excluded because of missing values. While there are some cities in the dataset as well, we exclude them from the analysis since NTLs are known to oversaturate in urban areas \cite{elvidge2013viirs} and do not allow a granular spatial analysis. Data on poverty rates in the SHRUG is based on the Socio-Economic Caste Census (SECC) of 2011 in India. Annual household consumption is predicted from information on asset ownership of individual households from the SECC based on the methodology developed in \cite{elbers2003micro}. The annual consumption mapping is then compared to the national poverty rate in 2012 which is below 2\$, or below Rs. 27.2, a day in rural areas. The resulting rate of households living in poverty in a village is described in column 1. The mean share of households living in poverty in our dataset is 0.35 although the standard deviation is high at 0.22. The median is close to the mean at 0.34 and the interquartile range indicates a symmetric distribution.

Electricity is a binary indicator which equals one if all households in a village have access to electricity for all purposes and zero otherwise. This is the case in 60\% of the villages. In the other 40 \%, either not all households have access to electricity or it is restricted for specific use cases only (domestic use, agricultural use). This kind of information, however, is only available for about 70\% of villages in the dataset. Hence, we restrict the dataset to this subset when an analysis involves electricity but use the full data when it does not.

Nighttime light information comes from the DMSP-OLS Stable Lights Composites which report the average annual NTL intensity on a scale from 0-63 in grid cells of about 1km $\times$ 1km. The resulting NTL indicator is a spatial average of these cells in a village. In Table 1, we only present summary statistics of NTLs for 2001 since values for 2011 are highly similar. NTL activity outside metropolitan areas is typically low which can be seen in the average of 3.38 across villages. The 25th percentile is still at 0 which implies that more than 25\% of villages in the dataset have virtually no detectable light footprint with DMSP-OLS.

Population counts (column (3)) from the census show that the average population per village is about 1360 although it has a large standard deviation of around 20,100. The distribution of the village population shows a right skew with a median of 776 far below the mean. Population values go into millions for high percentiles which underlines a somewhat arbitrary distinction between villages and cities for some cases in the Indian census that has been pointed out before \cite{asher2019socioeconomic}. 

On average, 18\% of the village population belongs to scheduled castes (SC) and around 16\% to scheduled tribes (ST). However, the empirical distribution of the two variables differs notably. While scheduled castes seem to be at least somewhat present in most villages, more than 50\% of villages have no members of scheduled tribes. This is not surprising since scheduled tribes are known to be concentrated mostly in eastern India where in some instances they account for more than 95\% of the state population. 

\begin{figure}[h]
  \centering
  \includegraphics[width=\linewidth]{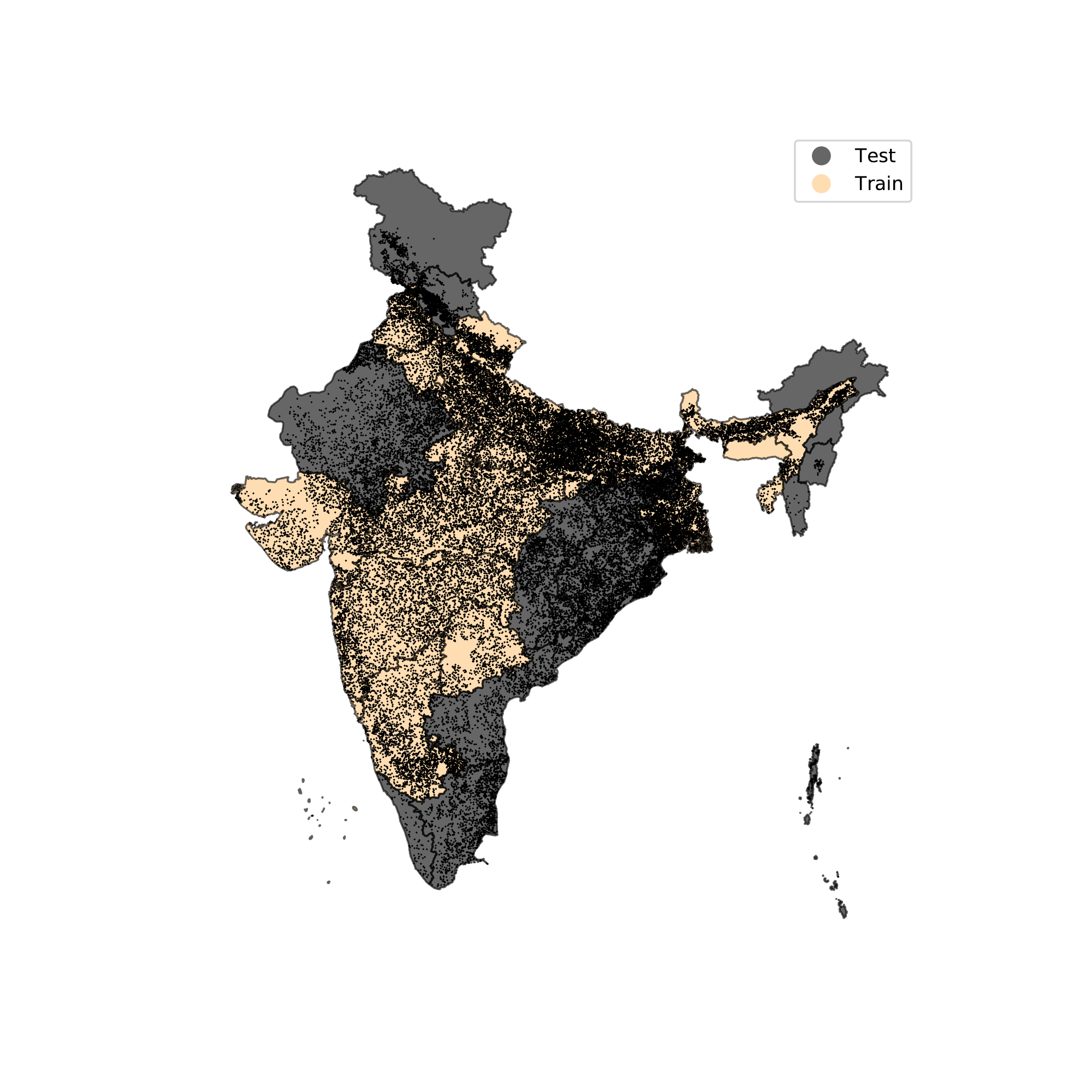}
  \caption{Spatial Training and Test Split by State}
  Yellow states belong to the training set, grey states to the test set. A spatial split is necessary to exclude overlap between training and test set. Village locations of a 10\% random sample are displayed as black points.
  \label{fig:SpatialSplit}
\end{figure} 

Although village locations are unique, some locations are very close and show similar characteristics. Hence, we spatially allocate villages of the same state together to the train and test portions. This makes the task of predicting indicators of development harder but prevents spatial label leakage between training and test set. To use the majority of our data for training, we randomly selected 20 states out of 36 (including union territories) for this. As population differs widely across states, states are drawn to the training set until a threshold of 2/3 of the population is crossed. Figure \ref{fig:SpatialSplit} displays the random spatial assignment of states to training and test set in more detail. Yellow states are the training set where the majority of villages is. In total, about 386,000 villages are in the training states which is around 73.6\% of the total number of observations. This is because the distribution of villages as well as population has spatial dependencies. Most of the highly populous states in central India are in the training set with the test states mostly located on the exterior of the map. Dark points in the graph show a 10\% random sample of village location in our dataset. We chose to display only a sample of villages for visibility purposes. 

 \section{Methods}
\subsection{Humanitarian Mapping with NTLs}
We restrict our focus on two widely used methods in humanitarian mapping with geospatial data. At first, we use a linear regression which is widely used in economics to map local development with NTLs \cite{bennett2017advances}. Typically, indicators of economic development are assumed to follow a lin-log relationship. Formally, this estimates: 
 \begin{equation}
    y_i = f(NTL_i) = \beta_0 + \beta_1 log(NTL_i) + \epsilon_i
    \label{eq:regressionbase}
\end{equation}

 where $y_i$ is the average development indicator of interest in village $i$, $NTL_i$ the average nightlight intensity and $\epsilon_i$ the residual of the prediction. In our case, $y_i \in $ \{Poverty,Electrification\}. To mitigate the effect of spatial dependencies and increase predictive accuracy we further include the coordinates of the respective village as input: 
\begin{equation}
    y_i = \beta_0 + \beta_1 log(NTL_i) + \beta_2 lat_i + \beta_3 lon_i + \epsilon_i
    \label{eq:regressionlatlon}
\end{equation}
 
 Second, we employ random forest regression which is a powerful tool to map the spatial distribution of humanitarian indicators with NTLs accurately \cite{rybnikova2021coloring,kondmann2021blinded}. Tuned hyperparameters include tree depth and the number of estimators in total. Hyperparameter selection is executed with a spatial three-fold cross validation over the training set. Similar to equation \ref{eq:regressionlatlon}, we fit: 
 \begin{equation}
    y_i = f(NTL_i,lat_i,lon_i)
    \label{eq:RF}
\end{equation}
 
 As random forests are invariant to the scaling of inputs, there is no necessity to transform the nightlights.
 
 \subsection{Fairness Audits}
Evaluating the fairness of algorithmic predictions requires a notion of what is understood as `fair' which can be difficult in practice. This is because a number of fairness criteria exist that are intuitively appealing but these criteria are only compatible with each other in unrealistic circumstances \cite{kleinberg2016inherent}. Causal approaches can help to resolve these conceptual roadblocks since they allow to formalize the meaning of what makes observations similar \cite{kilbertus2017avoiding}. Particularly, Kusner et al. \cite{kusner2017counterfactual} introduce the notion of counterfactual fairness which is the main criterion we use in this study. They impose that an algorithm should make the same decision for an observation in a community of interest in the real world and its true counterfactual if the same observation was outside the group. 
 
Establishing counterfactual fairness with certainty in this real-world setting is difficult since we can not rely on true counterfactuals of villages. This would require observing all villages twice, once with and once without high scheduled caste or tribe membership and comparing their nightlight activity.
We can, however, approximate this here by creating artificial counterfactuals by matching villages across groups. Our approach is inspired by propensity score matching \cite{rosenbaum1983central}, a widely used technique in econometrics to identify the causal effect of a treatment. In our case, we frame belonging to the community of interest such as a scheduled tribe as the treatment and try to identify the causal effect of this treatment on systematic prediction errors. 

For each village with an above-median share of scheduled caste or tribe membership ($s$) we find a counterfactual village ($\neg s$) from the below-median group. The matching algorithm finds the nearest neighbor from the ($\neg s$) villages group with replacement based on the following variables:
\begin{enumerate}
    \item Latitude
    \item Longitude
    \item Poverty
    \item Electrification
    \item Population 
\end{enumerate}

This allows us to directly compare the NTL activity in villages with and without scheduled caste or tribe membership but virtually identical location, poverty, electrification, and population. Let $\overline{\epsilon}_j=\sum_1^J (\hat{y_j}-y_j)$ be the average residual of the prediction of y for group j where $j \in \{s,\neg s\}$. Then, an algorithm shows an indication for bias if we can establish with statistical significance that $\overline{\epsilon}_s \neq \overline{\epsilon}_{\neg s}$. As $y_j$ is identical in both groups by design, this essentially tests if the prediction differs in both groups conditional on location, poverty, electrification, and population. In other words, if $\overline{\epsilon}_s \neq \overline{\epsilon}_{\neg s}$ villages in the group of interest would be targeted more or less frequently than villages in the comparison group resulting in unjust policy. 

Note that $\overline{\epsilon}_s \neq 0$ is not sufficient to establish prejudice since this only shows that a method over- or underpredicts development for villages of interest compared to all other villages. For example, say an algorithm predicts that comparably poor villages of a marginalized community are on average much wealthier than they actually are. While this is still not desirable, this might not necessarily show bias. The reason could also be an algorithm that generally overpredicts villages in poor subsets of the dataset and makes the same mistake for comparable villages outside the marginalized community. To evaluate fairness, these villages have to be compared to a designated comparison group such as the counterfactuals above. We test the statistical significance of the difference in means between the two groups with a t-test. Results with non-parametric Mann-Whitney U tests are omitted for the sake of brevity since they are virtually identical to the t-tests. 

 \section{Results}
\subsection{Humanitarian Mapping with NTLs}
Before analyzing fairness, a necessary step is to verify that the methods for mapping poverty and electricity with NTLs indeed work on our data. Table \ref{tab:Mapping} presents the results of mapping the development indicators from NTLs. The Table reports the share of explained variance $R^{2}$ for the two methods and inputs respectively on the test set. In column (1), we map village-level poverty rates with a linear regression based on NTLs only. With one input only, Linear Regression (LR) and Random Forest (RF) perform similarly with an $R^{2}$ of 11-12\%. When including coordinates, this improves to about 19\% with LR and 32\% with RF on the test set. Naturally, RF can utilize the additional information with more flexibility and achieves better results than LR.

\begin{table}[!htbp]
\centering
\caption{Mapping Development Indicators with Nightlights}
\begin{threeparttable}
\footnotesize
\setlength{\tabcolsep}{1.3\tabcolsep}
\input{prediction}\textbf{}
\end{threeparttable}
\label{tab:Mapping}
\end{table}

Since performance seems notably better when coordinates are directly included, we only estimate electricity including coordinates and henceforth focus on predictions based on NTLs and coordinates as input. For electricity, RF seems to overfit slightly as the less flexible LR models electricity rates more accurately with 36\% vs 25\% in $R^2$. Compared to previous studies of economic activity and nightlights, $R^2$ in Table \ref{tab:Mapping} are in general slightly lower than ranges in previous studies \cite{kondmann2021blinded,jean2016combining,weidmann2017using,bruederle2018nighttime}. 

This is not particularly surprising since the size of villages here is often even lower than the 10km*10km spatial resolution that is typically used in these studies. Predictive accuracy could be notably improved by geographical aggregation e.g. to the sub-district or district level or by including additional NTL characteristics beyond a spatial mean. Further, since we focus on rural villages in contrast to previous attempts there is comparably little heterogeneity in NTLs which may make it harder to distinguish economic development from space. Nevertheless, NTLs show strong associations with electricity and poverty in this dataset which confirms previous results \cite{asher2019socioeconomic}. After establishing a strong link between NTLs and development indicators, we proceed to estimate these indicators in the test set to analyze potential bias in the predictions. 

 \subsection{Fairness Audits}
\begin{table*}[!htbp]
\centering
\caption{Residual Correlation Relative to Matched Villages}
\begin{threeparttable}
\footnotesize
\setlength{\tabcolsep}{1.3\tabcolsep}
\input{matching}\textbf{}
\end{threeparttable}
\label{tab:Matching}
\end{table*}

Detailed results of the matching estimator are displayed in Table \ref{tab:Matching}. Results in columns (1) and (2) separate groups by share of scheduled tribes, and (3) and (4) for scheduled castes respectively. The coefficient is the mean difference in the prediction residual between the villages of interest and their matched counterfactuals $\overline{\epsilon}_s - \overline{\epsilon}_{\neg s}$. Panel 1 and 2 compare the same groups of villages with the only difference being that predictions in Panel 1 were obtained with LR and RF was used in Panel 2. All predictions are based on NTLs as well as village coordinates as inputs.

The prediction for poverty in villages in the group of scheduled tribes is on average 1.09 percentage points higher than in the comparison group. This difference is highly significant with a p-value below 0.001. This implies that poverty for scheduled tribes may be overestimated by using NTLs in a linear regression relative to comparable villages. This could, for example, be the case because similar villages without marginalized communities are favored with public investments in infrastructure and electricity which drive NTLs. This looks plausible in light of previous results around regional favoritism in India \cite{hodler2014regional}. Further, it would be in line with results from column (2) since predicted electrification rates are relatively about 1.3 percentage points lower for scheduled tribes. This result is again highly significant with a t-statistic close to 50 which underlines a clear difference in outcomes for both groups. Differences of around or even over 1 percentage point are not only significant but also notable in magnitude given an average poverty rate of around 35\%. 

Interestingly, the effect goes in the opposite direction for scheduled castes for both, poverty and electricity with LR. In the case of poverty (column 3), the coefficient of scheduled castes implies a predicted poverty rate that is on average 0.3 percentage points lower compared to the counterfactual group. This implies that targeting efforts based on NTLs may underestimate poverty in respective villages and, for example, these villages may receive less aid than needed as a consequence. This difference is once again highly significant but smaller in magnitude compared to scheduled tribes. Electricity is relatively overestimated for this group with an average, highly significant difference of 0.27 percentage points which is consistent with column (3). 

Panel 2 reinforces the robustness of findings from Panel 1. For all 4 specifications, we find a highly significant difference with the same directions. Effect magnitude varies slightly as it marginally increases for poverty and decreases for electricity. As poverty is estimated as a negative function of NTLs and electricity as a positive function of NTLs, it is not surprising that coefficients move in opposite directions. Panel 2 underlines that the problem is not easily solvable by a more flexible algorithm. The issue persists in our case which points to the fact that prejudice here may arise from a data bias rather than being induced by the algorithm. This indicates that seemingly exogenous geospatial sensor data may still reflect discriminatory dynamics on the ground. 

In summary, Table \ref{tab:Matching} provides an empirical indication for prejudice issues in humanitarian mapping with NTLs in India. These issues arise because villages of similar socio-economic conditions with differences in marginalized communities also show differences in light usage. We can not conclusively answer if this is the result of discrimination in the provision of public funds, differences in culture, or other reasons. Independent of the reasons, our algorithms implicitly treat villages of certain communities differently which is a major issue for humanitarian mapping applications in research and policy.

\section{Discussion}
We notice two key differences in our results for scheduled castes and scheduled tribes: 
\begin{enumerate}
    \item Smaller effect sizes for scheduled castes
    \item Opposite effect directions
\end{enumerate}
The smaller effect magnitude could be related to the centered distribution of scheduled caste members across villages. Since most villages will have some but not all members from scheduled castes, a potential discriminatory effect will be less visible compared to scheduled tribes. On the other hand, scheduled tribes are largely concentrated in villages of their community with relatively few outside members. 
Second, the different spatial distribution of scheduled castes and tribes may also provide insights into the opposite effect directions. Scheduled caste are more evenly distributed across all villages and economic predictions of them may hence be influenced by spatial overglow of NTLs. This is especially a known problem in populous regions where DMSP sensors are not precise enough and NTLs are too bright to separate the activity of individual cells from adjacent cells \cite{elvidge2013viirs}. So potentially wealthy and populous neighborhoods of scheduled castes may artificially increase their NTL activity and lead the algorithm to falsely predict higher living standards. For scheduled tribes, the geographic homophily may prevent such overglow effects and algorithms assume low development since NTL emissions are relatively lower given their living standards. 

While we restrict our focus on scheduled tribes and castes in this analysis, they are only one of many communities prone to be potentially discriminated against of interest in the context of Indian policy. We plan to investigate other indicators of interest such as religion, gender or age in future research since our framework is agnostic to the group of interest and the variable to be predicted. More generally, Table \ref{tab:Matching} tells a cautionary tale on potentially neglected biases in mapping human development with NTLs. It is ex ante unclear if this issue generalizes to other geographic regions or other geospatial data sources, too. It seems, however, plausible that such issues could arise in other contexts as well. 

This is because it is well known that machine learning algorithms may replicate or even amplify bias in the input data they receive \cite{mehrabi2019survey}. As long as input data such as NTLs are the result of complex data-generating processes on the ground, the potential for bias can hardly be excluded. This seems more obvious with data sources directly resulting from human behavior such as social media data or open street maps. Our work shows, however, that also seemingly exogenous sensor data can fall victim to these issues because they merely capture behavioral patterns on the ground. Therefore, our results underline the necessity for thoroughly understanding who we are capturing accurately in humanitarian mapping from space - and also who we are missing out on. 

\section{Conclusion}
In this paper, we study if geospatial predictions of human development based on nighttime lights with machine learning may suffer from fairness issues. For this purpose, we combine economic and census data for over 500.000 Indian villages with their geographic coordinates. We map rural poverty and electrification rates with NTLs accurately and proceed to analyze systematic prediction errors and their association with scheduled caste and tribe population. To allow for a valid comparison, we create synthetic counterfactual villages by matching villages from inside and outside the marginalized communities based on location and a variety of socio-economic indicators. Our framework allows us to approximate counterfactual fairness in the spatial predictions across villages and adds to the methodological toolset of causal approaches to fairness. 

Our results outline that villages with a high share of scheduled tribe population receive significantly higher poverty and lower electricity predictions compared to their comparable counterfactuals. This effect may be the result of a relative lack of public infrastructure and investment in these villages which is a key driver of NTLs. Conversely, we find that a high share of scheduled caste population corresponds to an underestimation of poverty relative to comparable villages outside of this community. This may cause policy-makers to neglect the scale of poverty issues in this marginalized community. These results are a warning sign for applications of NTLs in policy and research and open up a larger discussion about the representativeness of geospatial data sources. Fair and just prediction results are of particular importance in many humanitarian mapping applications that directly inform policy-makers. Our results underline that sensor data is not immune to representation issues and a thorough investigation with different algorithms, methods and locations is necessary to understand the scale of this issue in other applications.

\begin{acks}
We thank Sam Asher, Tobias Lunt, Ryu Matsuura, and Paul Novosad for providing and maintaining the SHRUG dataset. Further, we are grateful to Ariadna Pregel who provided excellent research assistance. 
\end{acks}

\bibliographystyle{ACM-Reference-Format}
\bibliography{sample-base}

\appendix

\end{document}

%% file: summary.tex
\begin{tabular}{lrrrrrr}
\toprule
{} &  Poverty Rate &  Electricity & NTLs &  Population &   Share SC &   Share ST \\
\midrule
N &            525255 &          370296 &        525255 &       525255 &  525255 &  525255 \\
Mean  &                 0.35 &               0.60 &             3.38 &         1359.03 &       0.18 &       0.16 \\
Std   &                 0.22 &               0.49 &             4.38 &        20098.74 &       0.20 &       0.31 \\
25\%   &                 0.18 &               0.00 &             0.00 &          364.00 &       0.01 &       0.00 \\
50\%   &                 0.34 &               1.00 &             2.65 &          776.00 &       0.12 &       0.00 \\
75\%   &                 0.50 &               1.00 &             4.83 &         1546.00 &       0.27 &       0.15 \\
\bottomrule
\end{tabular}

%% file: prediction.tex
\begin{tabular}{lccc}
\toprule
&\multicolumn{2}{c}{Poverty}&\multicolumn{1}{c}{Electricity}\\
\cmidrule(lr){2-3} 

    &\multicolumn{1}{c}{\(R^{2}\)}&\multicolumn{1}{c}{\(R^{2}\)}&\multicolumn{1}{c}{\(R^{2}\)}
    \\
\multicolumn{4}{l}{Methods} \\
\midrule

LR     & 0.1180      &   0.1877            & 0.3649      \\
RF     & 0.1130        &   0.3196                    & 0.2480   \\
\\
\multicolumn{4}{l}{Inputs} \\
\midrule
NTL     & (\checkmark)     &   (\checkmark)                   & (\checkmark)    \\
Coordinates     &        &   (\checkmark)                  &   (\checkmark)   \\

\bottomrule
\multicolumn{3}{l}{\footnotesize All Scores obtained on the Test Set.} \\
\end{tabular}

%% file: matching.tex
\begin{tabular}{lcccc}
\toprule
&\multicolumn{1}{c}{$\epsilon$(Poverty)}&\multicolumn{1}{c}{$\epsilon$(Electricity)}&\multicolumn{1}{c}{$\epsilon$(Poverty)}&\multicolumn{1}{c}{$\epsilon$(Electricity)}\\
     &  (1)   &  (2) & (3) & (4)    \\
\\
\multicolumn{5}{l}{Panel 1: Linear Regression} \\
\midrule
Scheduled Tribes & 0.0109*** & -0.0127***  && 
\\

Scheduled Castes     &&&  -0.0030***                         &  0.0027***      \\
\midrule

|t-stat|    &  18.7983 & 48.9393  & 42.8808 &      3.9104 \\
p & 0.000  & 0.000  & 0.000 &      0.000 \\
\\

\multicolumn{5}{l}{Panel 2: Random Forest} \\
\midrule
Scheduled Tribes     & 0.0121***                    & -0.0090***     & &     \\
Scheduled Castes     &&& -0.0040***                     & 0.0014***   \\
\midrule
|t-stat|    &  27.3220 & 18.2680  & 80.5898 &       65.4688 \\
p & 0.000  & 0.000  & 0.000 &      0.000 \\

\\

\midrule
N     & 78990      & 96272 &78990&      96272 \\
NTL     &(\checkmark) &   (\checkmark)     & (\checkmark)          & (\checkmark)  \\
Coordinates         &(\checkmark) &   (\checkmark)     & (\checkmark)          & (\checkmark)  \\

\bottomrule
\multicolumn{5}{l}{\footnotesize Standard errors in parentheses. * \(p<0.1\), ** \(p<0.05\), *** \(p<0.01\).} \\
\end{tabular}